\documentstyle[11pt,newpasp,twoside,epsf]{article}
\def\edcomment#1{\iffalse\marginpar{\raggedright\sl#1\/}\else\relax\fi}
\reversemarginpar

\begin{document}
\title{Evolution of X-ray Luminous Clusters from z=0.3 to z=0.8}
 \author{Isabella Gioia}
\affil{Istituto di Radioastronomia-CNR, Bologna, ITALY}
\affil{Institute for Astronomy, Honolulu, Hawaii, USA}

\begin{abstract}
Observations of the highest redshift clusters provide the longest lever
arm in the attempt to evaluate the evolution of their  bulk properties.
I present the most recent results on galaxy cluster evolution based on data
from the ROSAT North Ecliptic Pole survey. At $z>0.3$ there is a 
deficit of  clusters with respect to the local
universe which is significant at $>$ 4.7$\sigma$. The evolution 
appears to begin at L$_{0.5-2.0} > 1.8\times10^{44}$ erg s$^{-1}$ 
in the NEP survey data and goes in the same direction as the 
original EMSS result.  At lower redshifts there is no evidence for
evolution,  a result in agreement with all the existing cluster surveys.
\end{abstract}
\section{Introduction}
The spatial abundance of distant clusters and its redshift evolution
constrain cosmological parameters. For cosmological studies, however,
sample completeness with well-defined selection functions is required.
One of the cleanest way to avoid sample contamination 
is the selection of high redshift clusters by means of their X-ray emission. 
X-ray surveys are sensitive enough to detect objects at 
redshifts of order unity (e.g. MS1054$-$03 at z$=$0.83, 
Gioia \& Luppino 1994; RXJ1716.6+6708 at z$=$0.81, Gioia et al. 1999;
RXJ0848.9+4452 at $z=1.25$, Rosati et al. 1999; Cl J1226.9+3332
at $z=0.89$, Ebeling et al. 2001).
At these high redshifts some evolution is expected in the 
look-back time that approaches half the age of the universe.

Today all the determinations of the cluster X-ray luminosity function
(XLF) derived from existing surveys are in agreement for low redshifts
($<0.3$) and for low 
luminosities (L$_{0.5-2.0}<2.8\times10^{44}$ erg s$^{-1}$) indicating
no evolution. There is no unanimous opinion yet regarding the most 
luminous and most distant clusters known, but results in favor of
evolution are accumulating.  The first evidence for
negative evolution came from the Extended Medium Sensitivity Survey (EMSS; 
Gioia et al. 1990a; Stocke et al. 1991; Maccacaro et al. 1994) and 
was  discussed in terms of the cluster XLF; the volume density of high
redshift, high luminosity clusters is smaller than for nearby objects
(Gioia et al. 1990b; Henry et al. 1992). 
In the past several results seemed at odds with the EMSS evolution' s claim.
However, these were made without remembering that only the high
luminosity end of the XLF was claimed to evolve, a regime where the
supposedly contradictory studies had no data.

In this contribution I present additional evidence regarding the
evolution  of the cluster population based on the ROSAT All-Sky Survey 
observations  around the North Ecliptic Pole region (NEP). 
I use H$_{0}=$~50 km s$^{-1}$ Mpc $^{-1}$ and q$_{0}=$~0.5 throughout.

\section{The ROSAT North Ecliptic Pole survey}
The ROSAT NEP survey covers a $9^{\circ} \times 9^{\circ}$ region of
the  deepest area of the ROSAT All-Sky Survey (RASS; Tr\"umper et
al. 1991; Voges et al. 1999) where the scan circles converge and the effective 
exposure  time approaches  $40$ ks. 
Details on the ROSAT NEP
survey and the X-ray data can be found in Henry et al. 2001 and Voges et 
al. 2001.
\begin{figure}
\caption{ROSAT color image of the NEP survey region. The colors mostly
encode the X-ray photon energy. Red and yellow
for 0.11--0.40 keV, green for 0.40--1.00 keV, blue and purple for
1.00--2.40 keV photons. White implies bright sources in the 0.40--1.00
and/or 1.50--2.40 keV bands. From Voges et al. 2001.}
\end{figure}
The main difference between the NEP survey and the existing X-ray 
serendipitous  cluster surveys is that the NEP survey is 
both deep (median flux limit is f$_{0.5-2.0}=7.8\times10^{-14}$ erg
cm$^{-2}$ s$^{-1}$) and covers a contiguous area of sky (see
Fig. 1). Thus the NEP ROSAT survey database can be used to
examine large-scale structure in the  distribution of clusters
and other classes of extragalactic X-ray sources (e.g., AGN).
A concentration of 21 groups and clusters was indeed found during the 
analysis of the NEP sources (see Mullis et al. 2001).  
A total of 445 X-ray sources were detected with flux determinations
$>4\sigma$ in the $0.1-2.4$ keV band using the RASS-II processing.
All but two sources in the survey have been spectroscopically identified 
and redshifts  have been measured for the  extragalactic
population.  A complete and  unbiased  sample of 64 galaxy clusters
has been extracted. Nineteen clusters have 
a redshift greater than 0.3 with the highest at z$=$0.81.
In the course of identifying sources from the X-ray NEP
survey, several serendipitous discoveries were made. I mention here
the beautifully extended cluster RXJ1716+67 (Fig. 2), one among the
most distant (z$=$0.81) X$-$ray selected clusters so far published  
with a large number of spectroscopically determined cluster member 
velocities (see Gioia et al. 1999).

\begin{figure}
\caption{Three-color image of RXJ 1716+67. The brightest central galaxy
is  located in the lower center of the image, with a long filament 
leading out of the cluster to the northeast. Thirthy-seven galaxies 
have measured spectroscopic redshifts.  From  Clowe et al. 1998.}
\end{figure}

\section{The NEP Cluster Number-Counts Relation}
The integral number counts of galaxy clusters are a well established means
to verify reliability of sky coverage and completeness of identifications.
In Fig. 3, in addition to the NEP data,  the observed
cumulative number counts derived  from the 160 deg$^{2}$ (Vikhlinin et
al.  1998), the BCS (Ebeling et al. 1997), the WARPS (Jones et al. 1998), the
S-SHARC (Burke et al. 2001), the RDCS (Rosati et al. 1998), and the RASS1-BS
(De Grandi et al. 1999) are  shown. The NEP cluster number counts
are in agreement  within the errors with all the other independent
determinations, giving us confidence that our sample is complete.

\begin{figure}
\plotfiddle{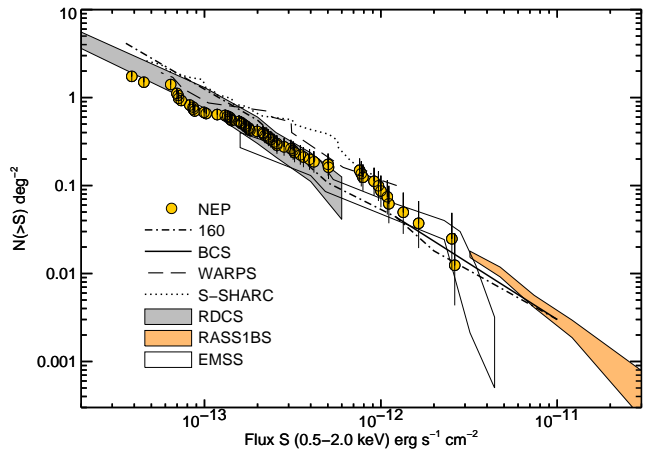}{4.9cm}{0}{50}{50}{-140}{-15}
\caption{Observed cumulative number counts from the NEP cluster survey
(circles) and other surveys. Adapted from Gioia et al. 2001.}
\end{figure}

\section{A deficit of high z and high L$_{X}$ clusters in the ROSAT
NEP}

The number of observed clusters in the NEP survey has been compared to the 
number of expected clusters, assuming no-evolution models.
The three local luminosity functions derived from the RASS1 (De Grandi
et al. 1999), REFLEX (B\"ohringer et al. 1998) and BCS (Ebeling et al.
1997) have been folded through the NEP sky coverage and then
integrated in the appropriate redshift and luminosity ranges.
The ranges of integration were $z<0.3$ and $0.3<z<0.85$ in redshift,
and $3\times10^{42}-10^{47}$ erg s$^{-1}$ in luminosity (0.5-2.0 keV).

For the $z<0.3$ redshift range, the number of clusters
expected from the three local samples and observed in the NEP are
consistent, with the the significance of difference equal to
$0.1-0.2\sigma$. For the $0.3<z<0.85$ range a value of 65.5
clusters is expected according to the RASS1,
a value of 55.9 according to the REFLEX  and
a  value of 44.2 according to the BCS. Only 19 NEP
clusters are observed in the same redshift and luminosity ranges.
The significance of deviation is 6.4$\sigma$, 7.2$\sigma$ or
4.7$\sigma$  depending on which of the three local XLF sample 
determinations is considered. There is a deficit  of clusters at 
L$_{0.5-2.0}>1.8\times10^{44}$ erg s$^{-1}$ and $z>0.3$ compared to
expectations from a non-evolving XLF (see Fig. 4 for a plot of the  
NEP XLF and three local XLFs.) This result goes in the
same  direction as the evolution derived from the EMSS survey.

\section{Discussion and Conclusions}
The ROSAT NEP survey detects clusters up to z$=$0.81. There is a
deficit  of clusters at high redshifts and high luminosity
compared to expectations from a non-evolving XLF. 
Five out of six surveys now find evolution. The original EMSS
(Gioia et al. 1990b; Henry et al. 1992), 160 deg$^{2}$ (Vikhlinin et 
al. 1998), SHARC  (Nichol et al. 1999), RDCS (Rosati et al. 2000) and 
NEP (Gioia et al. 2001) surveys are reporting
negative evolution at varying levels of significance from
$\sim$ 1$\sigma$ to greater than 5$\sigma$. The only X-ray survey
that does not find negative evolution is the WARPS (Jones et al. 2000;
Ebeling et al. 2001).
Evidence is  thus accumulating in favor of evolution at the high 
luminosity end of the XLF at high redshift.
The essentially complete identification rate of the
NEP survey gives us confidence that the deficit of clusters seen 
is not due to the fact that clusters have been missed.
Until larger and better characterized samples are available,
the results presented in this contribution represent the state of 
the art with respect to evolution of high redshift, high X-ray luminosity
clusters.
\begin{figure}
\plottwo{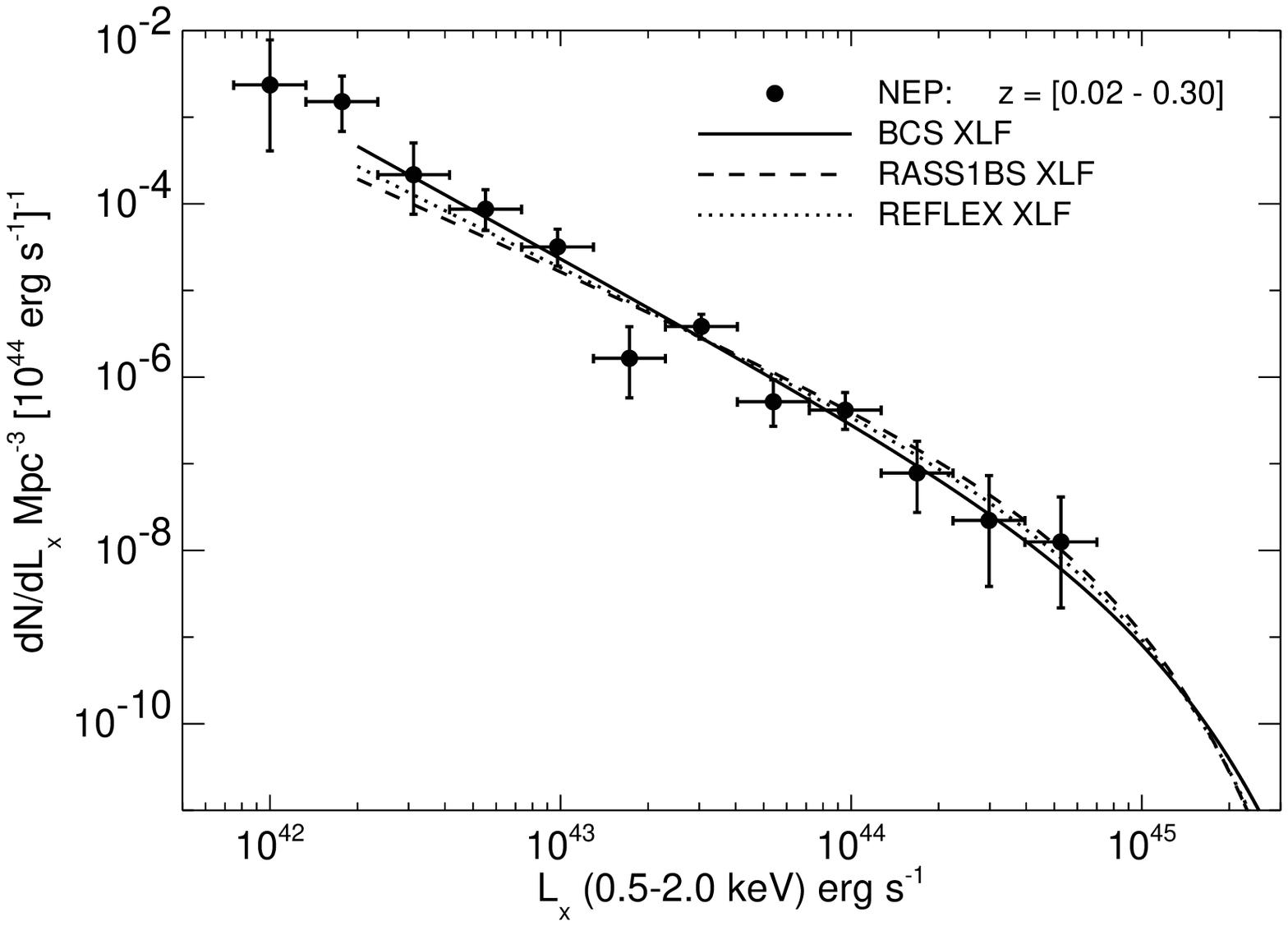}{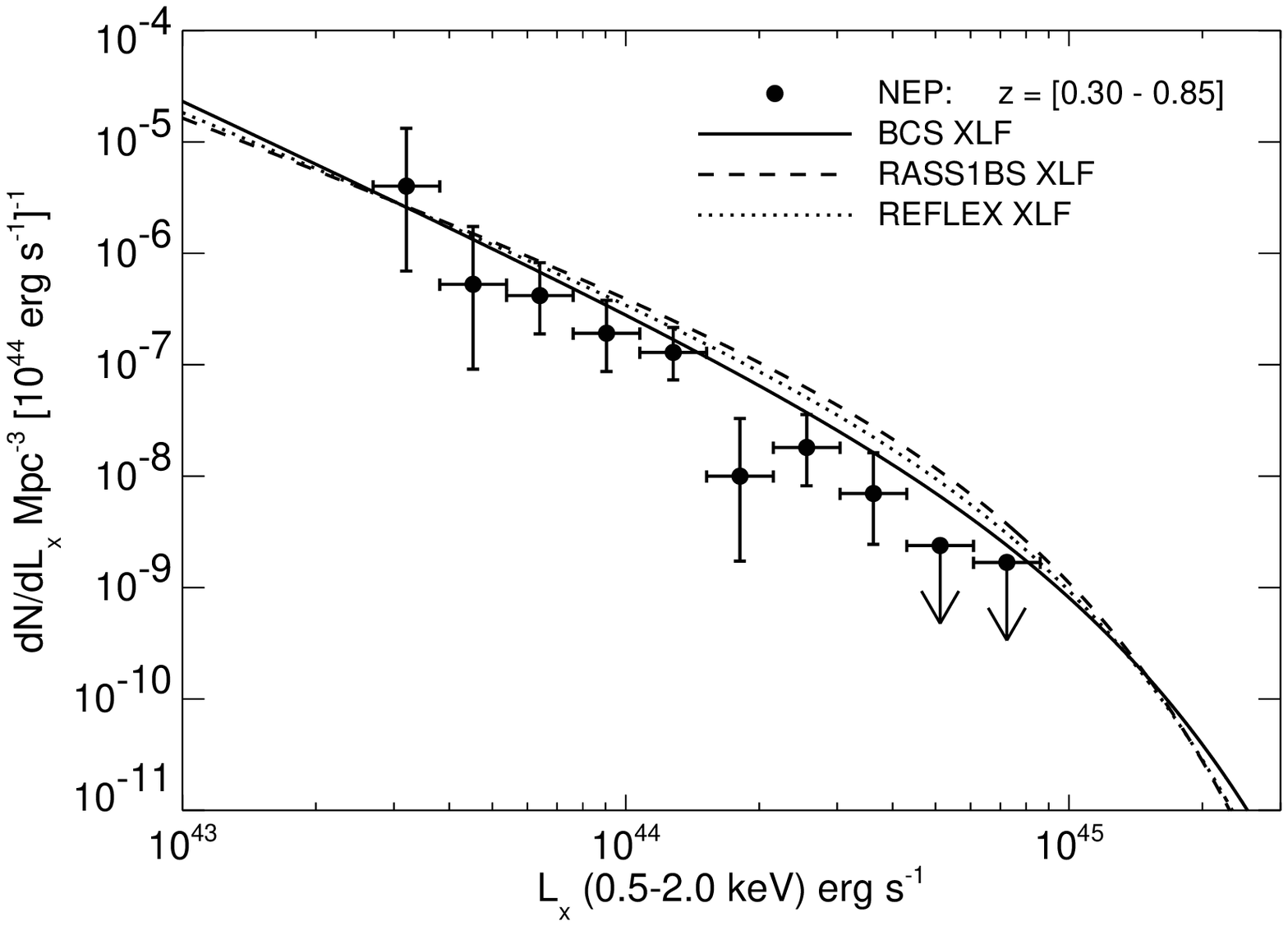}
\caption{The X-ray luminosity function for the NEP clusters in the
range $0.02<z<0.3$ (left panel) and for clusters in the range $0.3<z<0.85$
(right panel). The local XLF curves of the RASS1 (dashed), REFLEX (dotted)
and BCS (solid) are overplotted. From Gioia et al. 2001.}
\end{figure}

{\bf Acknowledgments} I wish to thank my collaborators who contributed 
to the success of the NEP survey: C. Mullis, P. Henry, H. B\"ohringer, U.
Briel, W. Voges and J. Huchra. The
ROSAT NEP survey was partially funded by NSF (AST91-19216
and AST95-00515), NASA (NAG 5-9994),  NATO (CRG91-0415) and from the
Italian Space Agency ASI.
Finally I  wish to thank Francesca Matteucci, Roberto Fusco-Femiano
and all the other membErs of the OC
for organizing a great meeting in a fascinating Italian island.

\end{document}